\documentstyle[prd,aps,preprint,epsf]{revtex}

\def\eps{\varepsilon}
\def\dot{\!\cdot\!}

\def\epf#1#2#3#4{\varepsilon_{#1#2#3#4}}
\tighten
\begin{document}
\draft
\title{\hfill {\small DOE-ER-40757-113} \\
\hfill {\small UTEXAS-HEP-98-10} \\
\hfill  {\small MSUHEP-80715} \\
\hfill \\ 
High energy photon-neutrino interactions}

\author{Ali Abbasabadi$^1$, Alberto Devoto$^2$, Duane A. Dicus$^3$ 
and Wayne W. Repko$^4$} 
\address{$^1$Department of Physical Sciences, Ferris State University, 
Big Rapids, Michigan 49307\\
$^2$Dipartmento di Fisica, Universit\'a di Cagliari and I.N.F.N., 
Sezione di Cagliari \\
$^3$Center for Particle Physics and Department of Physics\\ 
University of Texas, Austin, Texas 78712 \\
$^4$Department of Physics and Astronomy, Michigan State University \\ 
East Lansing, Michigan 48824}
\date{\today}
\maketitle

\begin{abstract}
A general decomposition of the amplitudes for the $2\to 2$ processes 
$\gamma\nu\to\gamma\nu$ and $\gamma\gamma\to\nu\bar{\nu}$ is obtained using 
gauge invariance and Bose symmetry. The restrictions implied by this 
decomposition are investigated for the reaction $\gamma\gamma\to\nu\bar{\nu}$ by
computing the one-loop helicity amplitudes in the standard model. In the center
of mass, where $\sqrt{s} = 2\omega$, the cross section grows roughly as
$\omega^6$ up to the threshold for $W$-boson production, $\sqrt{s} = 2m_W$.
Astrophysical implications of very high energy photon-neutrino interactions are
discussed.
\end{abstract}

\section{Introduction} \label{sec:1}

Investigations of the astrophysical importance of the $2\to 2$ processes 
$\gamma\nu\to\gamma\nu$, $\gamma\gamma\to\nu\bar{\nu}$
and $\nu\bar{\nu}\to\gamma\gamma$ have a long history. These reactions have been
examined using the four-Fermi interaction \cite{cm}, charged vector boson
theories \cite{mjl,ls}, the standard model \cite{cy,dr93} and model-independent 
parameterizations \cite{liu}. Because of the vector-axial vector nature of the
weak coupling, the cross sections are much smaller than a simple counting of
powers of the weak coupling, $G_F$, and the fine structure constant, $\alpha$,
in the diagrams of Fig.\,1 would suggest \cite{yang,gell}. For massless 
neutrinos, the cross sections vanish to order $G_F^2\alpha^2\omega^2$ 
\cite{gell}, and are known to be of order $G_F^4\omega^6$ when $\omega < m_e$ 
\cite{ls,dr93}. This is due to the fact that the scale of the loop integrals for
the diagrams in Fig.\,1 is set by the $W$-boson mass, $m_W$, rather than the
electron mass. As a consequence, for center of mass energies $2\omega$ between 1
keV and 1 MeV, the cross sections for $2\to 3$ processes such as
$\gamma\nu\to\gamma\gamma\nu$ are larger than the $2\to 2$ cross sections
\cite{dr97}.

Although of little practical importance in stellar processes where the energy
scale is $\lesssim$1 MeV, the $2\to 2$ processes could be important in very
high energy reactions such as $\nu\bar{\nu}\to\gamma\gamma$ \cite{wax},
particularly if the $\omega^6$ behavior persists for $\omega > m_e$. 
The next Section contains the development of a general decomposition of the 
elastic amplitude ${\cal A}(s,t,u)$ for $\gamma\nu\to\gamma\nu$. 
In Section\,\ref{sec:3}, we apply
this decomposition to the crossed channel $\gamma\gamma\to\nu\bar{\nu}$ and
obtain numerical results for the complete one-loop helicity amplitudes. This is
followed by a discussion of our results. 

\section{General decomposition of the $\bbox{\gamma\nu}$ elastic amplitude}
\label{sec:2}

The amplitude for the process $\gamma\nu\to\gamma\nu$ can be
expressed  as
\begin{equation}\label{amp}
{\cal A}(s,t,u) = \;\bar{u}(p_2)\gamma_{\mu}(1 + \gamma_5)u(p_1){\cal
M}_{\mu\alpha\beta}\,(\eps_1)_{\alpha}(\eps^*_2)_{\beta}\;,
\end{equation}
where ${\cal M}_{\mu\alpha\beta}$ is a sum of gauge invariant, Bose symmetric
tensors constructed from the neutrino momenta $p_1,\,p_2$ and the photon momenta
$k_1,\,k_2$. Construction of these tensors is aided by the fact that, for 
massless neutrinos, we have
\begin{eqnarray}
(p_1)_{\mu}\bar{u}(p_2)\gamma_{\mu}(1 + \gamma_5)u(p_1) & = &\;0\;, \\
(p_2)_{\mu}\bar{u}(p_2)\gamma_{\mu}(1 + \gamma_5)u(p_1) & = &\;0\;, \\
(k_1 - k_2)_{\mu}\bar{u}(p_2)\gamma_{\mu}(1 + \gamma_5)u(p_1) & = &\;0\;.
\end{eqnarray}
This suggests the use of the combinations
\begin{mathletters}
\begin{eqnarray}
p & = &\;p_1 + p_2\;, \\
k_+ & = &\;k_1 + k_2\;, \\
k_- & = &\;k_1 - k_2\;. 
\end{eqnarray}
\end{mathletters}
\noindent From the above, of $p_{\alpha},p_{\beta}\;\mbox{\rm and}\,p_{\mu}$, 
only $p_{\alpha}\,\mbox{\rm and}\,p_{\beta}$ will result in
non--vanishing contributions. Thus, the vector--vector and axial vector--axial
vector contributions to ${\cal M}_{\mu\alpha\beta}$ will consist of combinations
of $p_{\alpha},\,p_{\beta},\,(k_+)_{\mu},\,(k_2)_{\alpha}\,\mbox{\rm and}\,(k_1)
_{\beta}$ as well as $\delta_{\mu\alpha},\,\delta_{\mu\beta}\,\mbox{\rm and}\,
\delta_{\alpha\beta}$. 

The gauge invariant combinations of $p_{\alpha}$ and $p_{\beta}$ are
\begin{equation} \label{gau1}
p_{\alpha} - \frac{k_+\dot p}{k_+^2}(k_2)_{\alpha}\;\;\mbox{\rm and}\;\;
p_{\beta} - \frac{k_+\dot p}{k_+^2}(k_1)_{\beta}\,,
\end{equation}
and we note that $k_+\dot p = 2k_1\dot p = 2k_2\dot p$ and $k_+^2 = 2k_1\dot
k_2$. Gauge invariant second rank tensors involving $\delta_{\mu\alpha}$ and
$\delta_{\mu\beta}$ are
\begin{eqnarray} 
\left(k_1\dot k_2\delta_{\mu\alpha} - (k_1)_{\mu}(k_2)_{\alpha}\right) 
& \;\;\mbox{\rm and}\;\; &
\left(k_1\dot k_2\delta_{\mu\beta} - (k_2)_{\mu}(k_1)_{\beta}\right)
\label{gau21}\,, \\ [6pt]
\left(k_1\dot p\,\delta_{\mu\alpha} - (k_1)_{\mu}p_{\alpha}\right) 
& \;\;\mbox{\rm and}\;\; &
\left(k_2\dot p\,\delta_{\mu\beta} - (k_2)_{\mu}p_{\beta}\right)\,,\label{gau22}
\end{eqnarray}
and $\delta_{\alpha\beta}$ appears in the form
\begin{equation}\label{gau23}
\left(k_1\dot k_2\delta_{\alpha\beta} - (k_2)_{\alpha}(k_1)_{\beta}\right)\,.
\end{equation}
Gauge invariant third rank tensors can then be constructed from products of 
$(k_+)_{\mu}$, the vectors in Eq.\,(\ref{gau1}), and the second rank tensors in 
Eqs.\,(\ref{gau21}-\ref{gau23}). In doing so, it should be noted that 
combinations such as
\begin{eqnarray}
(k_1\dot k_2\delta_{\mu\alpha} - (k_1)_{\mu}(k_2)_{\alpha})(p_{\beta} -
\frac{k_+\dot p}{k_+^2}(k_1)_{\beta})\,, & & \label{gau31} \\ [6pt]
(k_1\dot p\,\delta_{\mu\alpha} - (k_1)_{\mu}p_{\alpha})(p_{\beta} -
\frac{k_+\dot p}{k_+^2}(k_1)_{\beta})\,, & & \label{gau32}\\ [6pt]
(k_+)_{\mu}(p_{\alpha} - \frac{k_+\dot p}{k_+^2}(k_2)_{\alpha})
(p_{\beta} - \frac{k_+\dot p}{k_+^2}(k_1)_{\beta})\,, & & \label{gau33}
\end{eqnarray}
are not linearly independent since 
\begin{equation}
\frac{k_+\dot p}{k_1\dot k_2}(\ref{gau31}) - (\ref{gau33}) = 2(\ref{gau32})\,.
\end{equation}
As a consequence, the tensors of Eq.\,(\ref{gau22}) can be omitted when 
constructing a complete set of gauge invariant third rank tensors.

The requirement of Bose symmetry can easily be included if one constructs gauge
invariant tensors that have a definite symmetry under the exchanges $k_1
\rightarrow -k_2$, $k_2 \rightarrow -k_1$ and $\alpha\leftrightarrow\beta$. The
vector $k_+ \rightarrow -k_+$ under these exchanges. It is possible to construct
four linearly independent, gauge invariant third rank tensors with definite 
symmetry under the exchange of the photons. They can be chosen to be
\begin{eqnarray}
T^{(1)}_{\mu\alpha\beta} & = & (k_1\dot k_2\delta_{\mu\alpha} - (k_1)_{\mu}
(k_2)_{\alpha})(p_{\beta} - \frac{k_+\dot p}{k_+^2}(k_1)_{\beta}) +
\nonumber \\ 
&   & (k_1\dot k_2\delta_{\mu\beta} - (k_2)_{\mu}(k_1)_{\beta})
(p_{\alpha} - \frac{k_+\dot p}{k_+^2}(k_2)_{\alpha})\,, \label{ten1}\\ [6pt]
T^{(2)}_{\mu\alpha\beta} & = & (k_1\dot k_2\delta_{\mu\alpha} - (k_1)_{\mu}
(k_2)_{\alpha})(p_{\beta} - \frac{k_+\dot p}{k_+^2}(k_1)_{\beta}) - 
\nonumber \\ 
&   & (k_1\dot k_2\delta_{\mu\beta} - (k_2)_{\mu}(k_1)_{\beta})
(p_{\alpha} - \frac{k_+\dot p}{k_+^2}(k_2)_{\alpha})\,, \label{ten2}\\ [6pt]
T^{(3)}_{\mu\alpha\beta} & = & (k_+)_{\mu}(k_1\dot k_2\delta_{\alpha\beta} -
(k_2)_{\alpha}(k_1)_{\beta})\,, \label{ten3}\\ [6pt]
T^{(4)}_{\mu\alpha\beta} & = & (k_+)_{\mu}(p_{\alpha} - \frac{k_+\dot
p}{k_+^2}(k_2)_{\alpha})(p_{\beta} - \frac{k_+\dot p}{k_+^2}
(k_1)_{\beta})\,.\label{ten4}
\end{eqnarray}
It can be seen that $T^{(1)}_{\mu\alpha\beta}$ is symmetric under the exchange
of photons, while $T^{(2)}_{\mu\alpha\beta}$, $T^{(3)}_{\mu\alpha\beta}$ and
$T^{(4)}_{\mu\alpha\beta}$ are antisymmetric.

To obtain tensors which correspond to the vector--axial vector interference
terms, one can use a set of independent momenta together with
$\epf\alpha\beta\lambda\rho$ to form gauge invariant tensors with definite
symmetry under the exchange of the photons. When this is done, one finds two
tensors with the appropriate properties. They have the form
\begin{eqnarray}
T^{(5)}_{\mu\alpha\beta} & = &(k_+)_{\mu}\epf\alpha\beta\lambda\rho (k_+)
_{\lambda}(k_-)_{\rho}\,, \label{gau41}\\ [6pt]
T^{(6)}_{\mu\alpha\beta} & = &(k_+)_{\mu}\epf\alpha\beta\lambda\rho (k_+)
_{\lambda}p_{\rho} - k_+^2\epf\mu\alpha\beta\lambda p_{\lambda} -
\delta_{\alpha\beta}\epf\mu\nu\lambda\rho (k_+)_{\nu}(k_-)_{\lambda}
p_{\rho} \nonumber \\
&  &+ p_{\alpha}\epf\mu\beta\lambda\rho (k_+)_{\lambda}(k_-)_{\rho} + 
p_{\beta}\epf\mu\alpha\lambda\rho (k_+)_{\lambda}(k_-)_{\rho}\,.\label{gau42}
\end{eqnarray}
The gauge invariance of Eq.\,(\ref{gau42}) can be verified using the identity
\begin{equation}
k_{\mu}\epf\alpha\beta\lambda\rho + k_{\alpha}\epf\beta\lambda\rho\mu +
k_{\beta}\epf\lambda\rho\mu\alpha + k_{\lambda}\epf\rho\mu\alpha\beta +
k_{\rho}\epf\mu\alpha\beta\lambda = 0\,,
\end{equation}
which holds for any vector $k$. Using this identity, the factor 
$\xi_{\mu} = \bar{u}(p_2)\gamma_{\mu}(1 + \gamma_5)u(p_1)$ in Eq.\,(\ref{amp}) 
can always be rewritten so that it is contracted with 
$\epf\mu\nu\alpha\beta$ instead of with $(k_+)_{\mu}$. The relation
\begin{equation}
-p_1\dot p_2\,\xi_{\mu} = \epf\mu\nu\alpha\beta \xi_{\nu}(p_1)_{\alpha}(p_2)_
{\beta}\,,
\end{equation}
then enables us to express any tensor containing an $\epf\mu\nu\alpha\beta$ as
a sum of scalar products. These are simply combinations of the $T^{(i)
}_{\mu\alpha\beta}$ of Eqs.\,(\ref{ten1}-\ref{ten4}). Explicitly, we find
\cite{foot1}
\begin{eqnarray}
T^{(5)}_{\mu\alpha\beta} & \propto &\,T^{(2)}_{\mu\alpha\beta}\,, \\
T^{(6)}_{\mu\alpha\beta} & \propto &\,\left(T^{(3)}_{\mu\alpha\beta} +
\frac{p\dot k_+}{k_+^2}T^{(1)}_{\mu\alpha\beta} - T^{(4)}_{\mu\alpha\beta}
\right)\,.
\end{eqnarray}
Consequently, the $T^{(i)}_{\mu\alpha\beta},\;i = 1,\cdots ,4$\,, 
are sufficient to parameterize the $\gamma\,\nu$ elastic
amplitude.

The tensor ${\cal M}_{\mu\alpha\beta}$ can thus be written
\begin{equation}
{\cal M}_{\mu\alpha\beta} = {\cal M}_1(s,t,u)T^{(1)}_{\mu\alpha\beta} + {\cal
M}_2(s,t,u)T^{(2)}_{\mu\alpha\beta} + {\cal M}_3(s,t,u)T^{(3)}_{\mu\alpha\beta}
+ {\cal M}_4(s,t,u)T^{(4)}_{\mu\alpha\beta}\,,
\end{equation}
where Bose symmetry requires
\begin{eqnarray}
{\cal M}_1(s,t,u) & = & {\cal M}_1(u,t,s)\,, \\
{\cal M}_j(s,t,u) & = &\,-{\cal M}_j(u,t,s),\,j = 2,3,4\,.
\end{eqnarray}
Using the center of mass frame, and the explicit forms of all the vectors, we 
can obtain the following helicity basis
\begin{eqnarray}
T^{(1)}_{\mu\alpha\beta}\,\xi_{\mu}(\eps_1)_{\alpha}(\eps^*_2)_{\beta} & = &
-s\cos(\theta/2)\biggl(t(\lambda_1 + 
\lambda_2 + 2\lambda_1\lambda_2) + 4s\lambda_1\lambda_2\biggr)\,, 
\label{hel1}\\ [6pt]
T^{(2)}_{\mu\alpha\beta}\,\xi_{\mu}(\eps_1)_{\alpha}(\eps^*_2)_{\beta} & = &
st\cos(\theta/2)(\lambda_1 - \lambda_2)\,, \label{hel2}\\ [6pt]
T^{(3)}_{\mu\alpha\beta}\,\xi_{\mu}(\eps_1)_{\alpha}(\eps^*_2)_{\beta} & = &
st\cos(\theta/2)(1 - \lambda_1\lambda_2)\,, \label{hel3}\\ [6pt]
T^{(4)}_{\mu\alpha\beta}\,\xi_{\mu}(\eps_1)_{\alpha}(\eps^*_2)_{\beta} & = &
-8\,\frac{s^2u}{t}\cos(\theta/2)\lambda_1\lambda_2\,,\label{hel4}
\end{eqnarray}
where $\theta$ is the angle between the incoming and outgoing neutrinos,
$\lambda_1$ is the helicity of the initial photon, $\lambda_2$ is the
helicity of the final photon and it is unnecessary to specify the neutrino
helicities since they have a definite handedness.
Notice that the first, third and fourth of these amplitudes are symmetric under
the exchange of $\lambda_1$ and $\lambda_2$, while the second is antisymmetric
under this exchange. Since time reversal invariance implies that the amplitude
should be symmetric under the exchange of initial and final helicities, the
second amplitude is T-violating. There are thus three helicity amplitudes if we
require time reversal symmetry. This is consistent with simple helicity counting
arguments. 

\section{The channel $\bbox{\gamma\gamma\to\nu\bar{\nu}}$} \label{sec:3}

The results of Section \ref{sec:2} can be expressed in the crossed channel
$\gamma\gamma\to\nu\bar{\nu}$ using the relations $p_1\to -p_1$ and $k_2\to
-k_2$, which imply the interchange $s\leftrightarrow t$. The combinations $k_+$
and $p$ become
\begin{mathletters}
\begin{eqnarray}
k_+ & = & k_1 + k_2 \to k_1 - k_2 \equiv k_- \\
p   & = & p_1 + p_2 \to -p_1 + p_2 \equiv -p
\end{eqnarray}
\end{mathletters}
These transformations allow us to write the cross channel counterparts of Eqs.\,
(\ref{ten1}-\ref{ten4}) as
\begin{eqnarray}
\tilde{T}^{(1)}_{\mu\alpha\beta} & = & (k_1\dot k_2\delta_{\mu\alpha} - 
(k_1)_{\mu}(k_2)_{\alpha})(p_{\beta} - \frac{k_-\dot p}{k_-^2}(k_1)_{\beta}) +
\nonumber \\ 
&   & (k_1\dot k_2\delta_{\mu\beta} - (k_2)_{\mu}(k_1)_{\beta})
(p_{\alpha} + \frac{k_-\dot p}{k_-^2}(k_2)_{\alpha})\,, \label{tten1}\\ [6pt]
\tilde{T}^{(2)}_{\mu\alpha\beta} & = & (k_1\dot k_2\delta_{\mu\alpha} - 
(k_1)_{\mu}(k_2)_{\alpha})(p_{\beta} - \frac{k_-\dot p}{k_-^2}(k_1)_{\beta}) - 
\nonumber \\ 
&   & (k_1\dot k_2\delta_{\mu\beta} - (k_2)_{\mu}(k_1)_{\beta})
(p_{\alpha} + \frac{k_-\dot p}{k_-^2}(k_2)_{\alpha})\,, \label{tten2}\\ [6pt]
\tilde{T}^{(3)}_{\mu\alpha\beta} & = & -(k_-)_{\mu}(k_1\dot k_2
\delta_{\alpha\beta} - (k_2)_{\alpha}(k_1)_{\beta})\,, \label{tten3}\\ [6pt]
\tilde{T}^{(4)}_{\mu\alpha\beta} & = & (k_-)_{\mu}(p_{\alpha} + \frac{k_-\dot
p}{k_-^2}(k_2)_{\alpha})(p_{\beta} - \frac{k_-\dot p}{k_-^2}
(k_1)_{\beta})\,.\label{tten4}
\end{eqnarray}
The amplitude can be expanded as
\begin{equation}
\tilde{{\cal M}}_{\mu\alpha\beta} = {\cal M}_1(t,s,u)\tilde{T}^{(1)}
_{\mu\alpha\beta} + {\cal M}_2(t,s,u)\tilde{T}^{(2)}_{\mu\alpha\beta} + 
{\cal M}_3(t,s,u)\tilde{T}^{(3)}_{\mu\alpha\beta}
+ {\cal M}_4(t,s,u)\tilde{T}^{(4)}_{\mu\alpha\beta}\,,
\end{equation}
where
\begin{eqnarray}
{\cal M}_1(t,s,u) & = & {\cal M}_1(u,s,t)\,, \label{sym1}\\ 
{\cal M}_j(t,s,u) & = &\,-{\cal M}_j(u,s,t),\,j = 2,3,4\,.\label{sym2}
\end{eqnarray}
In this case, the center of mass helicity basis is
\begin{eqnarray}
\tilde{T}^{(1)}_{\mu\alpha\beta}\,\tilde{\xi}_{\mu}(\eps_1)_{\alpha}
(\eps_2)_{\beta} 
& = &\frac{1}{2}s\sin\theta\biggl(s(\lambda_1 - \lambda_2 + 2\lambda_1
\lambda_2) + 4t\lambda_1\lambda_2\biggr)\,, \label{hhel1}\\ [6pt]
\tilde{T}^{(2)}_{\mu\alpha\beta}\,\tilde{\xi}_{\mu}(\eps_1)_{\alpha}
(\eps_2)_{\beta} 
& = &-\frac{1}{2}s^2\sin\theta(\lambda_1 + \lambda_2)\,, \label{hhel2}\\ [6pt]
\tilde{T}^{(3)}_{\mu\alpha\beta}\,\tilde{\xi}_{\mu}(\eps_1)_{\alpha}
(\eps_2)_{\beta} 
& = &\frac{1}{2}s^2\sin\theta(1 + \lambda_1\lambda_2)\,, \label{hhel3}\\ [6pt]
\tilde{T}^{(4)}_{\mu\alpha\beta}\,\tilde{\xi}_{\mu}(\eps_1)_{\alpha}
(\eps_2)_{\beta} 
& = & 4tu\sin\theta\lambda_1\lambda_2\,.\label{hhel4}
\end{eqnarray}
Here, $\tilde{\xi}_{\mu} = \bar{u}(p_1)\gamma_{\mu}(1 + \gamma_5)v(p_2)$ and
$\theta$ is the angle between photon 1, which is in the $z$ direction, and
the outgoing neutrino. Using the helicity basis, we can express the
helicity amplitudes $\tilde{{\cal A}}_{\lambda_1\lambda_2}(s,z)$, where $z =
\cos\theta$, as
\begin{eqnarray} \label{hamps}
\tilde{{\cal A}}_{++}(s,z) & = &\sin\theta\left[s(t - u){\cal M}_1(t,s,u)
+ s^2{\cal M}_3(t,s,u) + 4tu{\cal M}_4(t,s,u)\right]\,,\label{hamp1} \\
\tilde{{\cal A}}_{+-}(s,z) & = &\sin\theta\left[-2st{\cal M}_1(t,s,u) - 
4tu{\cal M}_4(t,s,u)\right]\,,\label{hamp2} \\
\tilde{{\cal A}}_{-+}(s,z) & = &\sin\theta\left[2su{\cal M}_1(t,s,u) -
4tu{\cal M}_4(t,s,u)\right]\,,\label{hamp3} \\
\tilde{{\cal A}}_{--}(s,z) & = &\sin\theta\left[s(t - u){\cal M}_1(t,s,u) + 
s^2{\cal M}_3(t,s,u) + 4tu{\cal M}_4(t,s,u)\right]\,, \label{hamp4}
\end{eqnarray}
where we have assumed time reversal symmetry and omitted ${\cal M}_2(t,s,u)$.
From Eqs.\,(\ref{hamp1}) and (\ref{hamp4}), is is clear that $\tilde{{\cal
A}}_{++}(s,z) = \tilde{{\cal A}}_{--}(s,z)$. Moreover, since $t =
-\case{1}{2}s(1 - z)$ and $u = -\case{1}{2}s(1 + z)$, the symmetries of Eqs.\,
(\ref{sym1}) and (\ref{sym2}) imply
\begin{equation} \label{nflip}
\tilde{{\cal A}}_{++}(s,z) = -\tilde{{\cal A}}_{++}(s,-z)\,.
\end{equation}
Similarly, we find that the helicity flip amplitudes satisfy
\begin{equation}\label{flip}
\tilde{{\cal A}}_{+-}(s,z) = -\tilde{{\cal A}}_{-+}(s,-z)\,.
\end{equation}

To explore these general results, we calculated the standard model diagrams
of Fig.\,1 using a nonlinear $R_{\xi}$ gauge such that the coupling between the
photon, the $W$-boson and the Goldstone boson vanishes \cite{gauge,pall,dr93}.
These diagrams can be decomposed in terms of scalar two-point, three-point and
four-point functions, which are expressible in terms of dilogarithms \cite{pv}. 
The expressions for the helicity amplitudes in terms of dilogarithms were 
evaluated numerically and the results checked against FORTRAN codes developed 
for one-loop integrals \cite{dk,ff}. In doing so, we assumed that $s >> 
m_{\ell}^2$, where $m_{\ell}$ is the lepton mass, and showed that the 
dependence on $\ln(m_{\ell}^2)$ or $\ln^2(m_{\ell}^2)$ from individual diagrams 
actually cancels.

The results for the helicity amplitudes are shown in Figs.(2) and (3), where the
cross sections for various helicities are plotted for $\sqrt{s} = 20$\,GeV and
$\sqrt{s} = 200$\,GeV using
\begin{equation}
\frac{d\sigma_{\lambda_1\lambda_2}}{dz} = \frac{1}{32\pi s}|\tilde{{\cal
A}}_{\lambda_1\lambda_2}|^2\,.
\end{equation}
These figures clearly illustrate the symmetries of Eqs.\,(\ref{nflip}) and
(\ref{flip}), particularly the vanishing of the non-flip amplitudes at $z = 0$.
The total cross section for unpolarized photons is given by
\begin{equation}
\sigma_{\gamma\gamma\to\nu\bar{\nu}} = \frac{1}{64\pi s}\int_{-1}^1dz\left[
|\tilde{{\cal A}}_{++}(s,z)|^2 + |\tilde{{\cal A}}_{+-}(s,z)|^2\right]\,,
\end{equation}
and is plotted in Fig.\,(4). This figure illustrates the roughly $s^3$ behavior
of the total cross section up to the threshold for $W$ pair production. Also
shown in dashes is the helicity non-flip contribution to the cross section,
which can be seen to be much smaller than the helicity flip
contribution. This feature does not appear to be related to any symmetry, but it
is reminiscent of the low energy case, where the non-flip amplitudes vanish
\cite{dr93}. 

\section{Discussion and conclusions}

As mentioned in the Introduction, the low energy $2\to 2$ photon-neutrino 
cross sections are much smaller than the corresponding $2\to 3$ cross sections
for center of mass energies between 1 keV and 1 MeV. Recently, the $2\to 3$
cross sections have been computed for a range of center of mass energies from
well below $2m_e$, where the cross sections vary as $\omega^{10}$, to well above
$2m_e$, but less than $m_W$ \cite{dkr98}. For $m_e << \omega << m_W$, the only 
scale in the $2\to 3$ processes is $m_W$ or, equivalently, $G_F$. From 
dimensional considerations, the center of mass cross section must behave as 
$G_F^2\,\omega^2$ in this range of energy. Explicitly, it is found that
$\sigma_{\gamma\gamma\to\nu\bar{\nu}\gamma}$ can be expressed as
\begin{equation}
\sigma_{\gamma\gamma\to\nu\bar{\nu}\gamma} = 5.68\times
10^{-14}\left(\frac{\omega}{m_e}\right)^2\,\mbox{\rm fb}\,.
\end{equation}
For the $2\to 2$ process $\gamma\gamma\to\nu\bar{\nu}$, we expect an $\omega^6$
behavior, and a fit to the data in Fig.\,(4) yields
\begin{equation}\label{sigggnn}
\sigma_{\gamma\gamma\to\nu\bar{\nu}} = 4.0\times 10^{-31}
\left(\frac{\omega}{m_e}\right)^6\,\mbox{\rm fb}\,.
\end{equation}
These expressions are equal for $\omega = 1.94\times 10^4m_e$ or 
$\sqrt{s}\sim 20$\,GeV. Thus, for sufficiently high energies, the $2\to 2$
process dominates the $2\to 3$ process.

To assess the importance of very high energy photon-neutrino interactions in
processes of interest in cosmology, consider the scattering of high energy
neutrinos from the cosmic neutrino background. This process has recently been
studied by assuming that the neutrino collision produces a $Z$ at resonance
whose decay chain contains photons and protons which could account for the flux
of $> 10^{20}$\,eV cosmic rays \cite{wax}. For the processes considered here, 
the $\nu\bar{\nu}$ collision would directly produce high energy photons. The
cross section for $\nu\bar{\nu}\to\gamma\gamma$ can be obtained from 
Eq.\,(\ref{sigggnn}) by supplying a factor of 2. Assuming the relic neutrinos 
have a small mass \cite{wax,mass}, the cross section in the rest frame of the 
target is
\begin{equation}
\sigma_{\nu\bar{\nu}\to\gamma\gamma} = 5.6\times
10^{-42}\left(\tilde{m_{\nu}}\tilde{E_{\nu}}\right)^3\;\mbox{\rm cm$^2$}\,,
\end{equation}
where $\tilde{m_{\nu}}$ is $m_{\nu}/1\,{\rm eV}$ and $\tilde{E_{\nu}}$
is $E_{\nu}/10^{21}\,{\rm eV}$. The interaction rate 
on the cosmic neutrino background is $\sigma_{\nu\bar{\nu}\to\gamma\gamma}
n_{\nu}c$, where $n_{\nu}$ is the relic neutrino density. If this is multiplied
by $t_0$, the age of the universe, the condition that at least one interaction
occur is
\begin{equation}
\sigma_{\nu\bar{\nu}\to\gamma\gamma}n_{\nu}ct_0  = 1\,,
\end{equation}
or
\begin{equation}
\tilde{m_{\nu}}\tilde{E_{\nu}}  =  \frac{5.6\times 10^{13}\,{\rm cm}^{-2/3}}
{(n_{\nu}ct_0)^{1/3}}\,.
\end{equation}
Taking $n_{\nu} = 56$\,cm$^{-3}$ and $t_0 = 15\times 10^9$\, years,
$\tilde{m_{\nu}}\tilde{E_{\nu}} = 6.07\times 10^3$, which translates into
$\sqrt{s} = 3.5$\,TeV. Even if there were sources of such high energy neutrinos
(which seems unlikely), this is well above the region for which the $s^3$
behavior is valid. In fact, the relation $(n_{\nu}ct_0)^{-1} = 1.26\times
10^{-30}\;\mbox{\rm cm$^2$}$ together with Fig.\,(4) show that
$\sigma_{\nu\bar{\nu}\to\gamma\gamma}$ is not large enough for any value of $s$
to attenuate a high energy neutrino flux given the present values of $n_{\nu}$
and the age of the universe.

The temperature at which the reaction $\nu\bar{\nu}\to\gamma\gamma$ ceases to
occur can be determined from the reaction rate per unit volume
\begin{equation} \label{rho}
\rho = \frac{1}{(2\pi)^6}\int\frac{d^3p_1}{e^{E_1/T} + 1}
\int\frac{d^3p_2}{e^{E_2/T} + 1}\sigma|\vec{v}|\,,
\end{equation}
where $\vec{p}_1$ and $\vec{p}_2$ are the neutrino and antineutrino momenta,
$E_1$ and $E_2$ their energies, $|\vec{v}|$ is the flux and $T$ 
the temperature. Using the invariance of $\sigma E_1E_2|\vec{v}|$,
the relationship between $\sigma |\vec{v}|$ in the center of mass 
frame and any other frame is 
\begin{equation}\label{sigv}
\sigma |\vec{v}| = \sigma_{CM}\frac{2E_{\rm CM}^2}{E_1E_2}\,.
\end{equation}
If the angle between $\vec{p}_1$ and $\vec{p}_2$ is $\theta_{12}$, 
Eq.\,(\ref{sigv}) gives
\begin{equation}
\sigma_{\nu\bar{\nu}\to\gamma\gamma}|\vec{v}| = 1.6\times
10^{-30}\frac{E_1^3E_2^3}{m_e^6}\sin^8(\theta_{12}/2)\;\mbox{\rm fb}\;.
\end{equation}
The integration is straightforward and Eq.\,(\ref{rho}) gives
\begin{equation}
\rho_{\nu\bar{\nu}\to\gamma\gamma} = \frac{1.6\times 10^{-30}\,\mbox{\rm
fb}}{5\pi^4}\left(\frac{31}{32}\Gamma(6)\zeta(6)\right)^2\frac{T^{12}}{m_e^6}\,,
\end{equation}
where $\zeta(x)$ is the Riemann zeta function. The interaction rate
$R_{\nu\bar{\nu}\to\gamma\gamma}$ can be obtained by dividing
$\rho_{\nu\bar{\nu}\to\gamma\gamma}$ by the neutrino density $n_{\nu} =
3\zeta(3)T^3/4\pi^2$, and we find
\begin{equation}
R_{\nu\bar{\nu}\to\gamma\gamma} = 7.3\times 10^{-24}T_{10}^9\;\mbox{s$^{-1}$}
\,,
\end{equation}
with $T_{10}$ denoting $T/10^{10}\,{\rm K}$. Expressing the
age of the universe as $t = 2T_{10}^{-2}$\,s, the condition for at least one
interaction, $R_{\nu\bar{\nu}\to\gamma\gamma}t\sim 1$ gives $T\sim 1.6$\,GeV,
which is within the region of validity of the $s^3$ behavior.

\acknowledgements

One of us (A. D.) wishes to thank the Department of Physics and Astronomy of
Michigan State University for its support and hospitality.
This research was supported in part by the U.S. Department of Energy under
Contract No. DE-FG013-93ER40757 and in part by the National Science Foundation 
under Grants No. PHY-93-07980 and PHY-98-02439.

\begin{figure}[h]
\hspace{1.5in}
\epsfysize=2.5in \epsffile{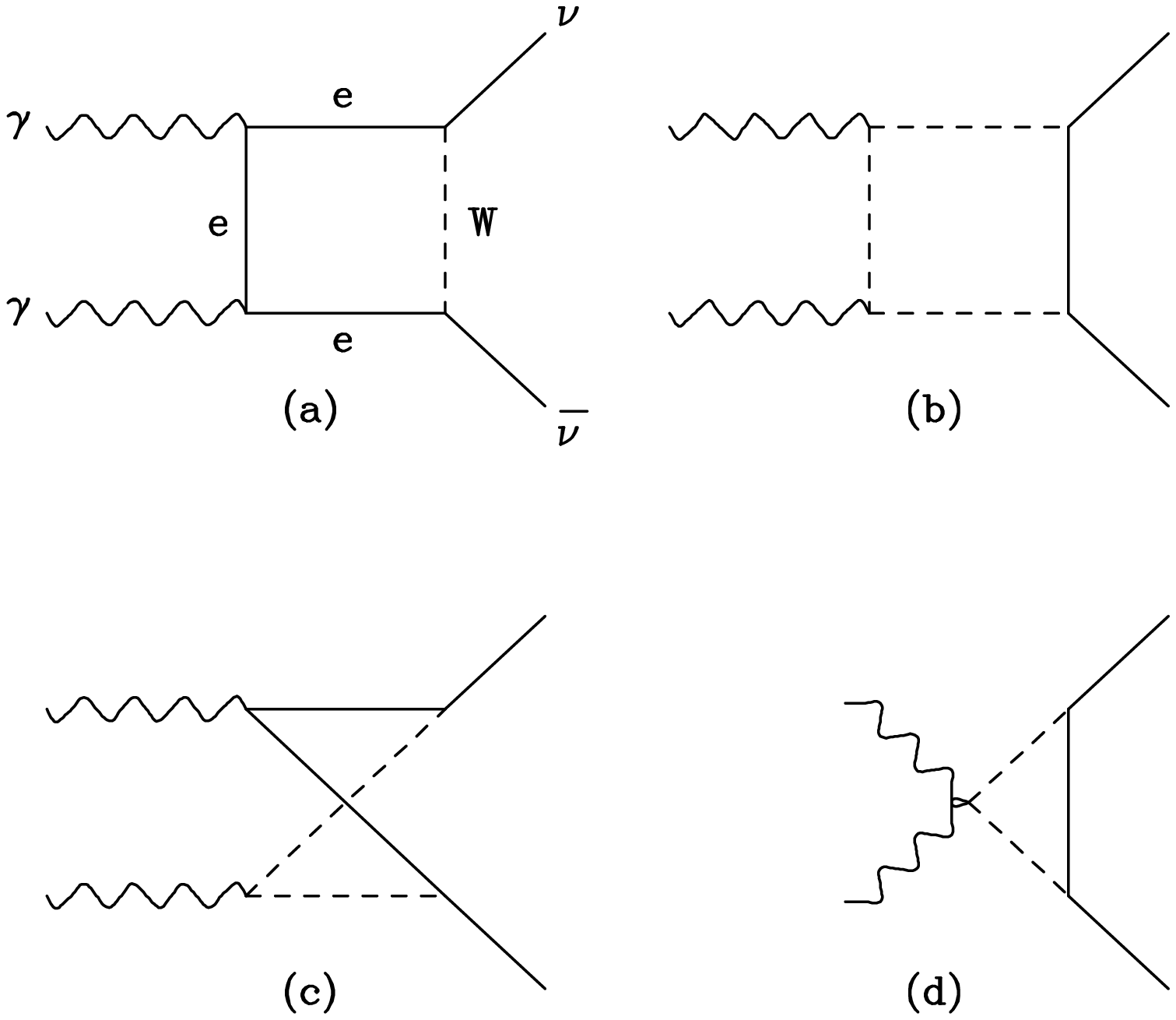} 
\vspace{10pt}
\caption{Diagrams for the process $\gamma\gamma\protect\to  \nu_e\bar{\nu}_e$. 
For each of (a), (b), (c) there is also a diagram with the photons 
interchanged.}
\end{figure}

\begin{figure}[h]
\hspace{1.2in}
\epsfysize=2.8in \epsffile{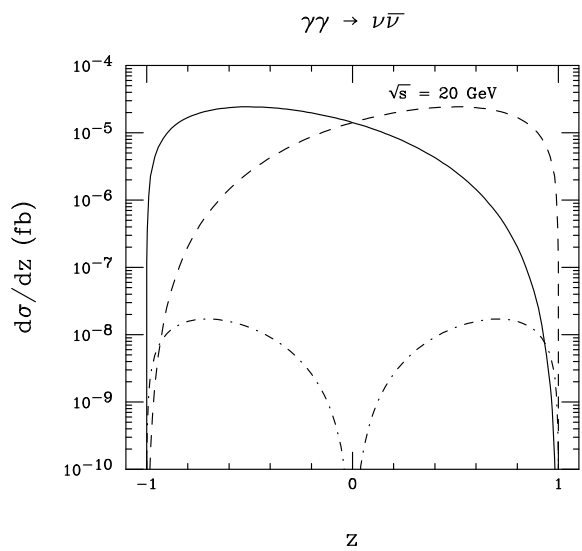} 
\vspace{10pt}
\caption{The helicity dependent differential cross sections for 
$\gamma\gamma\protect\to\nu\bar{\nu}$ are shown for $\protect\sqrt{s} = 20$
\,GeV. The
solid line is $d\sigma_{+-}/dz$, the dashed line is $d\sigma_{-+}/dz$ and the
dot-dashed line is $d\sigma_{++}/dz$.}
\end{figure}

\newpage

\begin{figure}[h]
\hspace{1.2in}
\epsfysize=2.8in \epsffile{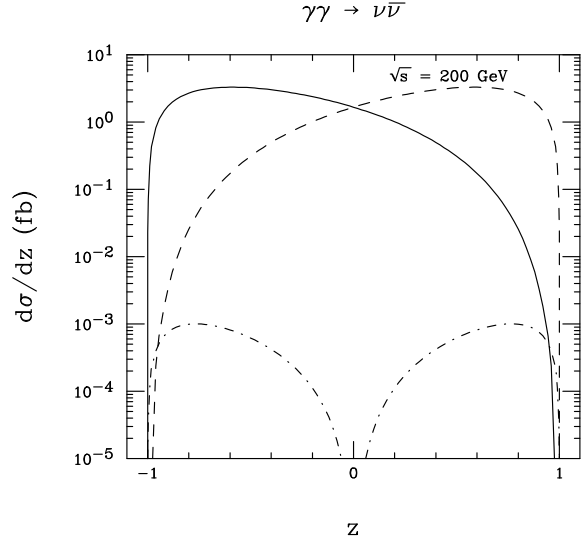} 
\vspace{10pt}
\caption{Same as Fig.\,(2) with $\protect\sqrt{s} = 200$\,GeV.}
\end{figure}

\begin{figure}[h]
\hspace{1.0in}
\epsfysize=2.8in \epsffile{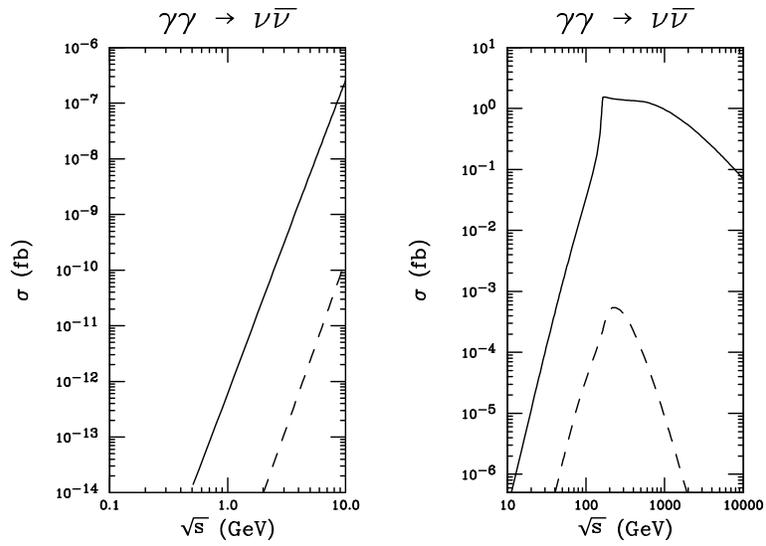}
\vspace{10pt}
\caption{The solid line is the total cross section
$\sigma_{\gamma\gamma\protect\to\nu\bar{\nu}}$ and the dashed line is the
contribution from the helicity non-flip amplitudes.}
\end{figure}


\begin{thebibliography}{99}
\bibitem{foot1} This is most easily shown using Eqs.\,(\ref{hel1}-\ref{hel4})
and the corresponding expressions for $T^{(5)}_{\mu\alpha\beta}$ and $T^{(6)}_
{\mu\alpha\beta}$\,.
\bibitem{mass} The corrections to the $s^3$ behavior of Eq.\,(\ref{sigggnn})
due to a small neutrino mass are of order $m_{\nu}^2/s$, which is negligible.
\end{thebibliography}
\end{document}